\documentclass[aip,jcp,amsmath,amssymb,floatfix,reprint,twocolumn]{revtex4-2}
\usepackage{color,soul}
\usepackage{graphicx}
\usepackage{amsmath, amsthm, amssymb,mathtools}
\usepackage{amsthm}
\usepackage{multirow}
\usepackage[normalem]{ulem}
\usepackage{soul}
\usepackage{hyperref,comment}

\usepackage{cancel}

\usepackage{multirow}
\usepackage{booktabs} 
\usepackage{array}

\usepackage{bm}
\usepackage{bigdelim}
\usepackage{array}
\usepackage{float}
\usepackage[table]{xcolor}
\arrayrulecolor{gray}
\newcommand{\mycomment}[1]{}

\newcommand{\be}{\begin{equation}}
\newcommand{\ee}{\end{equation}}
\newcommand{\bea}{\begin{eqnarray}}
\newcommand{\eea}{\end{eqnarray}}
\makeatletter

\newcommand{\Rmnum}[1]{\expandafter\@slowromancap\romannumeral #1@}
\makeatother\usepackage{array, makecell}

\begin{document}

\title{Speeding up Brownian escape via intermediate finite potential barriers}

\author{Vishwajeet Kumar}
\affiliation{The Institute of Mathematical Sciences, C.I.T. Campus, Taramani, Chennai 600113, India}
\affiliation{Homi Bhabha National Institute, Training School Complex, Anushakti Nagar, Mumbai 400094, India}
\author{Ohad Shpielberg}
\affiliation{Haifa Research Center for Theoretical Physics and Astrophysics, University of Haifa, Abba Khoushy Ave 199, Haifa 3498838, Israel}
\author{Arnab Pal} 
\email{arnabpal@imsc.res.in}
\affiliation{The Institute of Mathematical Sciences, C.I.T. Campus, Taramani, Chennai 600113, India}
\affiliation{Homi Bhabha National Institute, Training School Complex, Anushakti Nagar, Mumbai 400094, India}


\begin{abstract}

The mean first-passage time (MFPT) for a Brownian particle to surmount a potential barrier of height $\Delta U$ is a fundamental quantity governing a wide array of physical and chemical processes. According to the Arrhenius Law, the MFPT typically grows exponentially with increasing barrier height, reflecting the rarity of thermally activated escape events. In this work, we demonstrate that the MFPT can be significantly reduced by reshaping the original single-barrier potential into a structured energy landscape comprising multiple intermediate barriers of lower heights, while keeping the total barrier height $\Delta U$ unchanged. Furthermore, this counterintuitive result holds across both linear and nonlinear potential profiles. Our findings suggest that tailoring the energy landscape---by introducing well-placed intermediate barriers---can serve as an effective control strategy to accelerate thermally activated transitions. These predictions are amenable to experimental validation using optical trapping techniques.
\end{abstract}

\maketitle

\begin{quotation}
\noindent
\textbf{Barrier crossing underlies many activation processes, such as chemical reactions and protein folding, where a key quantity is the crossing time, governing transition rates like product formation in chemical reactions. By modeling these transitions as a Brownian particle moving in an external potential, we uncover a surprising mechanism to accelerate them. We show that adding intermediate barriers within the original potential landscape, while keeping the overall barrier height unchanged, can reduce the average crossing time. Intriguingly, the introduction of more intermediate barriers yields even faster transitions. These theoretical predictions are amenable to experimental validation using optical trapping techniques.
}
\end{quotation}

\noindent

\section{Introduction}
\label{intro}
Thermally activated escape processes are widespread in science, playing a central role in phenomena such as chemical reactions, protein folding, and nucleation during crystallization. These processes typically involve a system or agent surmounting an energy barrier to transition into a more stable or desired state. In chemical reactions, reactant molecules must overcome an activation barrier to break existing bonds and form new ones, thereby enabling the transformation of reactants into more stable products. In another instance, during protein folding, a polypeptide chain can undergo a series of conformational transitions across multiple intermediate states, by overcoming energy barriers before ultimately attaining its native, stable structure. A key quantity of interest across these scenarios is the average time required to cross the barrier(s). For instance, in chemical reactions, the reaction rate can be estimated as the inverse of this average transition time. In proteins, the folding times characterize the time it takes a protein to attain functional conformation, which is essential for executing its biological activities.
Gaining insight into the factors that affect this time, which can be referred as the first-passage time, is thus crucial and a challenging study spanning physics, chemistry, biology and other interdisciplinary studies — see \cite{elber2020molecular,metzler2014first,redner2001guide,szabo1980first,bray2013persistence,zhang2016first,berezhkovskii2021distributions} for a review on this topic.

A foundational theoretical framework for understanding barrier-crossing times was established by Kramers in his landmark work \cite{kramers1940brownian}. He showed that a Brownian particle attempting to overcome a high potential barrier provides an effective model for calculating reaction rates in chemical processes. To illustrate this, imagine a Brownian particle confined in a metastable state of a potential landscape $U(x)$, while in contact with a thermal reservoir at temperature $T$. To escape the trap, the particle must overcome a barrier of height $\Delta U$ defined by the potential $U(x)$. The crossing time depends on the interplay between two key forces: the deterministic force arising from the potential $U(x)$, and a stochastic force resulting from thermal fluctuations—random collisions with the bath particles. Kramers showed that in the limit where the barrier height $\Delta U$ is much larger than the thermal energy $k_B T$ (with $k_B$ being the Boltzmann constant), the mean escape time $\mathcal{T}$ takes on a well-defined form
\begin{equation}
\label{eq: Kramers equation}
    \mathcal{T}=A e^{\Delta U/k_BT},
\end{equation}
where the exponential factor is universal, depending solely on the barrier height $\Delta U$ and not on the particular shape of the potential, while the pre-exponential factor $A$ is sensitive to the specific form of the potential and can also depend on the nature of dynamics \cite{gardiner2009stochastic}. 

\begin{figure*}
    \centering
    \includegraphics[width=0.8\linewidth]{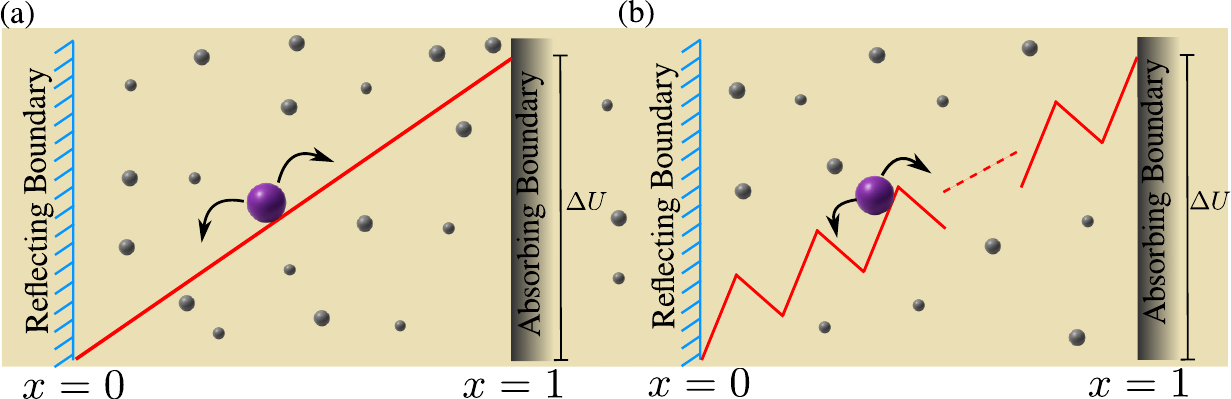}
    \caption{Panel (a): Schematic of a Brownian particle (large sphere) that starts at $x=0$, which is a reflecting boundary, and is eventually absorbed at $x=1$, which is the absorbing boundary, after overcoming a potential barrier of height $\Delta U$. 
    By altering the potential structure by introducing intermediate potential barriers as in panel (b), we would like to examine whether the mean first-passage time to reach the absorbing boundary can be reduced than that from the unaltered potential configuration as in panel (a). The Brownian particle is assumed to be in contact with a thermal bath surrounded by the bath particles (indicated by the smaller spheres).}
    \label{fig:main schematic}
\end{figure*}

Despite the passage of decades since its inception, Kramers’ theory of barrier crossing continues to be a cornerstone in the study of reaction dynamics and remains an active field of research \cite{hanggi1990reaction}. Modern investigations have extended the classical framework into new domains, motivated by developments in nonequilibrium statistical mechanics, complex systems, and active matter. Recent studies have explored a wide range of phenomena: barrier crossing in systems driven by active noise \cite{woillez2019activated, militaru2021escape, caprini2019active, caprini2021correlated}, escape dynamics of composite particles such as dimers \cite{asfaw2012exploring, lyngdoh2024escape, singh2017kramers}, and the influence of long-range temporal correlations or memory effects on the escape time statistics \cite{barbier2024long}. Others have focused on many-body systems where particle interactions play a pivotal role in modifying the escape dynamics \cite{kumar2024arrhenius, kumar2024emerging,kumar2024inferring}.

A prominent theme within this body of work is the search for mechanisms that can \textit{accelerate} the escape process—critical in applications ranging from chemical kinetics to molecular machines. Several innovative strategies have emerged to reduce the mean barrier crossing time. These include: Noise engineering -- modulating the memory time of colored noise has been shown to significantly influence the crossing dynamics \cite{kappler2018memory}; Stochastic resetting -- introducing resetting protocols—where particles are intermittently returned to a specific location—can lead to an optimal escape rate under certain conditions \cite{pal2024random,pal2024channel,singh2020resetting,ahmad2019first,capala2023optimization,ray2020diffusion}; Time-dependent driving: Applying weak, time-periodic forces to the system can enhance the likelihood of barrier crossing via stochastic resonance mechanisms \cite{wellens2003stochastic}; Removal of single maximum trapping time along the path of a tracer in disordered systems \cite{holl2023controls}. Collectively, these studies hint toward a paradigm shift in our understanding of thermally activated escape, expanding Kramers' classical picture into a richer and more diverse landscape of nonequilibrium strategies for controlling barrier crossing times.

Despite significant progress, research on how to manipulate potential landscapes to control the mean barrier crossing time—or equivalently, the mean first-passage time (MFPT)—remains an active and evolving field. This is also important since diffusion of particles, ions, molecules, or even living microorganisms in confined geometries such as tubes and channels plays a key role across various scales in natural and technological processes \cite{karger2003diffusion,gershow2007recapturing,dagdug2024diffusion}. One promising direction involves reshaping the potential profile, such as subdividing or optimizing the energy barrier to facilitate more efficient escape \cite{wagner1999intermediates,  palyulin2012finite,chupeau2020optimizing}. 
A notable initial contribution in this area is due to Wagner and Kiefhaber~\cite{wagner1999intermediates}, who numerically studied the effect of partially folded intermediates along the reaction pathway on the rates of protein folding. By modeling the folding process as the diffusion of reaction coordinates in a potential energy landscape, they showed that placing intermediates within the barrier, without changing its overall height, can enhance folding rates if the intermediate barriers are of moderate heights. Another important contribution in this direction was made by Palyulin and Metzler\cite{palyulin2012finite}, who demonstrated that replacing a simple linear potential with a piecewise linear potential consisting of two segments (while maintaining a fixed barrier height $\Delta U$) allows for tuning the segment slopes to minimize the MFPT. It was shown that the optimal MFPT achieved in this segmented configuration was lower than that of the original linear potential. Another crucial development along this line was due to Chupeau et al. \cite{chupeau2020optimizing}, who experimentally demonstrated that the MFPT of the Brownian particle for reaching a target can be reduced by guiding it through a suitably engineered potential barrier profile.
This acceleration of the first-passage process was observed in both the underdamped and overdamped regimes.

This naturally leads to the following compelling questions: Can a faster completion be achieved when a potential is composed of multiple segments instead of just two? Can the systematic control of the potential shape—achieved through structural modifications of individual segments—indeed expedite the completion process? The purpose of this work is to study a comprehensive model combined with theory and simulations to address such questions. By leveraging the knowledge of such arrangements, it may become possible to design potential landscapes that serve as regulators of transport, where  modifications to the potential shape can systematically control the rate at which large molecules, ions, or metabolites escape through confining channels.

To investigate these questions, we consider a prototypical set-up of a Brownian particle diffusing in a landscape depicted in Fig. \ref{fig:main schematic}(a); the particle starts at position $x=0$, which acts as a reflecting boundary while the boundary at $x=1$ is absorbing. Let $\mathcal{T}$ denote the mean first-passage time (MFPT) for the particle to reach $x=1$ by crossing the potential barrier of height $\Delta U$. Now, consider the alternative setup shown in Fig. \ref{fig:main schematic}(b), where the potential is replaced with a piecewise structure by introducing multiple intermediate barriers of smaller heights—creating several linear segments—while keeping the overall barrier height $\Delta U$ unchanged. We denote the MFPT in this modified potential by $\mathcal{T}^\text{mod}$. The central question we explore is whether a suitably chosen piecewise linear potential can yield a lower MFPT, such that $\mathcal{T}^\text{mod} < \mathcal{T}$? Using analytical tools from first-passage theory, we show that by carefully tuning the slopes of the individual segments in the piecewise linear potential, the MFPT can be significantly reduced. Moreover, we find that increasing the number of intermediate barriers—i.e., the number of adjustable segments—further decreases the MFPT, yielding a powerful control for optimizing escape times. These insights are not limited to linear potentials. We extend our analysis to non-linear landscapes as well, demonstrating that analogous improvements in MFPT can be achieved by replacing smooth non-linear potentials with carefully constructed piecewise non-linear profiles, as illustrated in Fig. \ref{fig:schematics multiple kinks L H}(b). 
These general findings establish the introduction of intermediate barriers as a general and robust technique for facilitating faster barrier crossing in diffusive systems.

The paper is organized in the following way. The system set up is described in section \ref{sec: set up formalism and kramers formula}. This section further discusses the first-passage formalism to compute the MFPT. 
 In section \ref{sec:single kink linear potential}, we review the case of single barrier \cite{palyulin2012finite} and extend to other potentials in section \ref{sec:single kink harmonic potential}. Section(\ref{sec:Multiple kinks}) discusses the impact of introducing multiple intermediate barriers in both linear and non-linear potential landscapes. It is shown that a systematic arrangement of barriers can lead to a successive reduction in the MFPT with increasing number of barriers employed. We conclude in Sec.~\ref{sec:conclusion} with a future outlook. Some additional details (including the effect of intermediate barriers to the fluctuations in first-passage time) are included to the appendices.

\section{Setup and Review of the first-passage Formalism}
\label{sec: set up formalism and kramers formula}
Consider a Brownian particle, whose position $x(t)$ is described by the overdamped Langevin equation in an external potential $U(x)$
\cite{zwanzig2001nonequilibrium,risken1996fokker}
\begin{equation}
\label{eq:Langevin eqn}
    \dot{x}(t)=-\frac{U'(x)}{\gamma}+\sqrt{2D}\eta(t),
\end{equation}
where $\eta(t)$ is a Gaussian white noise with the following statistical properties
\begin{equation}
    \langle \eta(t)\rangle=0, \qquad \langle \eta(t)\eta(t')\rangle=\delta(t-t').
\end{equation}
By the fluctuation dissipation theorem, $D=k_BT/\gamma$, where $\gamma$ is the viscosity of the surrounding medium, which we assume to be unity, implying $D=k_BT$. 
We confine the particle in the region $x\in[0,1]$ as shown in Fig. \ref{fig:main schematic}(a). The particle starts from $x_0\in[0,1]$ and it must surmount a potential barrier of height $\Delta U=U(x=1)-U(x=0)$, in order to be absorbed at $x=1$. Furthermore, without loss of generality, we set $U(x=0)=0$ which implies that $U(x=1)=\Delta U$.

In this specific set up, the MFPT, $\mathcal{T}(x_0)$, is the average time it takes for the particle to get absorbed at $x=1$ starting from an initial position $x_0\in[0,1]$. To find $\mathcal{T}(x_0)$, we start by writing the backward Fokker Planck equation, for the survival probability $\mathcal{P}(t,x_0)$, which denotes the probability that the particle has not been absorbed up to time $t$, given that it started at position $x_0$ at $t = 0$. From $\mathcal{P}(t,x_0)$, the first-passage time density is derived using
\begin{equation}
    \mathcal{F}(t,x_0)=-\frac{\partial}{\partial t}\mathcal{P}(t,x_0),
\end{equation}
where $\mathcal{F}(t,x_0)dt$ represents the probability that the particle has been absorbed in the time between $t$ and $t+dt$. The MFPT is then obtained using the following relation
\begin{equation}
    \mathcal{T}(x_0)=\int_0^\infty  t \, \mathcal{F}(t,x_0)dt,
\end{equation}
which satisfies the following backward Fokker-Planck equation
\cite{gardiner2009stochastic}:
\begin{equation}\label{mfpt differential equation}
    -U'(x_0)\frac{\partial}{\partial x_0}\mathcal{T}(x_0)+k_BT\frac{\partial^2}{\partial x_0^2}\mathcal{T}(x_0)=-1,
\end{equation}
where note that the initial position $x_0$ is now being treated as a variable \cite{gardiner2009stochastic}.
To determine $\mathcal{T}(x_0)$, this equation must be solved subject to the two boundary conditions
\begin{equation}
\label{Eq:bdy condition}
    \partial_{x_0}\mathcal{T}(x_0)\Big|_{x_0=0}=0,\quad  \text{and} \qquad \mathcal{T}(x_0=1)=0.
\end{equation}

 To compute the corresponding MFPT for an arbitrary potential, sometimes it is useful to recast Eq.(\ref{mfpt differential equation}) into an integral form \cite{gardiner2009stochastic}
\begin{equation}
\label{mfpt integrated eqn}
    \mathcal{T}(x_0)=\frac{1}{k_BT}\int_{x_0} ^1 dx'' e^{U(x'')/k_BT}\int_{ 0} ^{x''}dx' e^{-U(x')/k_BT},
\end{equation}
where we have already imposed the boundary conditions. Setting $x_0=0$, we will henceforth denote the MFPT as $\mathcal{T}$. Moreover, we will also adopt the following notation: $U_N^\text{L}$ refers to a potential with $N$ barriers/kinks, where each segment is linear. The associated MFPT for this potential is denoted by $\mathcal{T}_N^\text{L}$. Similarly, $U_N^\text{H}$ represents a potential with $N$ barriers/kinks, where each segment is harmonic. The corresponding MFPT for this potential is denoted by $\mathcal{T}_N^\text{H}$.

\section{Introducing one intermediate barrier}
\label{sec:single kink}
In this section, we demonstrate that introducing an intermediate barrier (or kink) in  potential landscapes and adjusting the strength of the segments on either side of the kink, while keeping the kink position and overall barrier height fixed, can reduce the MFPT compared to the original potential profile.

In Section \ref{sec:single kink linear potential}, we briefly reproduce the results of Palyulin and Metzler\cite{palyulin2012finite} for the case of a linear potential. In Section \ref{sec:single kink harmonic potential}, we extend it to consider a harmonic potential, investigating whether such modification results in a reduction of the MFPT in this case as well.

\subsection{MFPT reduction in linear potential}
\label{sec:single kink linear potential}
Consider a linear potential described as follows:
\begin{equation}
\label{eq:no kink linear potential}
    \frac{U^\text{L}_0(x)}{\Delta U}=x, \qquad \qquad 0\leq x\leq 1
\end{equation}

\begin{figure}
    \centering
    \includegraphics[width=1\linewidth]{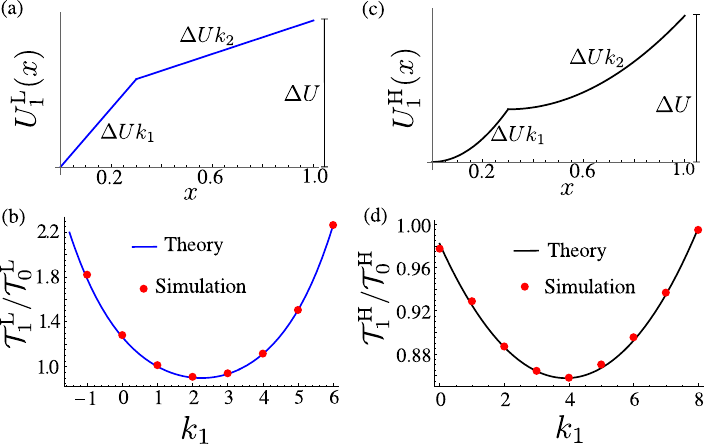}
    \caption{Panels (a) $\&$ (c): Piecewise linear and harmonic potential with a single kink at $x=a=0.3$. Panel (b): The inverse speed up factor, $\mathcal{T}_1^\text{L}/\mathcal{T}_0^\text{L}$, for linear case is plotted as a function of $k_1$ with $\Delta U = 3$ and $a=0.3$. For a range of $k_1$ values, $\mathcal{T}_1^\text{L}/\mathcal{T}_0^\text{L}$ is less than $1$, implying that the modification has accelerated the process completion in these regimes. Panel (d): The inverse speed-up factor, $\mathcal{T}_1^\text{H}/\mathcal{T}_0^\text{H}$, for harmonic case is plotted as a function of $k_1$ for $\Delta U = 3$ and $a=0.3$. Similar to the linear case, $\mathcal{T}_1^\text{H}/\mathcal{T}_0^\text{H}<1$ for several $k_1$ values.
  Simulation data are shown as points, and the analytical solution is plotted as continuous curve and $k_BT$ is taken to be $1$.
  }
    \label{fig:single kink linear and non linear}
\end{figure}

\begin{figure}
    \centering
    \includegraphics[width=1\linewidth]{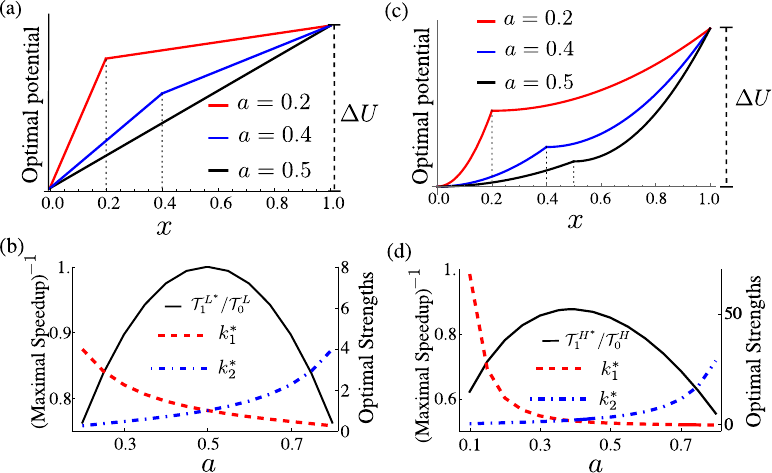}
    \caption{ Panels (a) $\&$ (c) show the optimal potential profiles (for various values of $a$) for the case of one kink in linear and harmonic potentials respectively.
Panel (b) shows the variation of the inverse maximal speed-up factor and the corresponding optimal strengths with $a$, for the linear case, with $\Delta U=3$. The left vertical axis represents this dimensionless factor, while the right vertical axis indicates the corresponding optimal strengths, $k_1^*$ and $k_2^*$. For $a=0.5$, we observe $k_1^*=k_2^*$, implying that the optimal potential is the original linear potential itself in this case from panel (a). 
Panel (d) depicts the variation of these quantities for harmonic case, for $\Delta U=3$. It shows that the inverse maximal speed-up factor never reaches unity. Similar to the linear case, there exists a range of $a$ values for which $k_1^* > k_2^*$, while for other values, $k_2^* > k_1^*$ as shown in panel (d).
 The MFPTs are determined analytically and the corresponding optimal values have been found numerically, while fixing $k_BT$ to unity.
}
    \label{fig:single kink minimal MFPT vs a}
\end{figure}

To determine the MFPT associated with this potential, we solve Eq.(\ref{mfpt differential equation}) subject to the two boundary conditions in Eq.(\ref{Eq:bdy condition}). This leads to the following expression for the MFPT for the particle starting at $x_0=0$:
\begin{equation}
\label{eq:asymptotics linear}
    \mathcal{T}_0^\text{L}=\frac{1}{\Delta U^2}(k_BTe^{\Delta U/k_BT}-\Delta U -k_BT).
\end{equation}
We now modify the linear potential into a piecewise-linear form consisting of two segments, defined as:
\begin{equation}
\label{Linear potential}
    \frac{U^\text{L}_1(x)}{\Delta U}=
    \begin{cases}
        k_1x, \qquad  \qquad \qquad \qquad 0\leq x\leq a \\
        k_2(x-a)+k_1a, \qquad \quad a\leq x\leq 1
    \end{cases}
\end{equation}
where, $k_1$ and $k_2$ can be thought of as the strength of the potential segments which can take any real values.
Imposing the setup constraint $U_1^\text{L}(x=1)=\Delta U$ implies 
\begin{equation}
\label{eq:fbc sk}
    \Delta U =\Delta U [k_1a+k_2(1-a)]
\end{equation}
Fig.\ref{fig:single kink linear and non linear}(a) depicts this potential (Eq.(\ref{eq:no kink linear potential})) for one particular values of $k_1$, $k_2$ and $a$.
To determine the associated MFPT, we solve Eq.~(\ref{mfpt differential equation}) separately over two regions: Region~I, defined on the interval $x_0 \in [0, a]$, and Region~II, defined on $x_0 \in [a, 1]$. The general solutions in each region contain constants of integration, which are determined by enforcing the two boundary conditions, together with the following matching conditions at the position of the kink (see Appendix~\ref{app:fifth} for details)
\begin{equation}
\label{eq: connecting condition 1}
    \mathcal{T}_{1,I}^\text{L}(x_0 = a) = \mathcal{T}_{1,II}^\text{L}(x_0 = a)
\end{equation}
and
\begin{equation}
\label{eq: connecting condition 2}
    \partial_{x_0} \mathcal{T}_{1,I}^\text{L}(x_0 = a) = \partial_{x_0} \mathcal{T}_{1,II}^\text{L}(x_0 = a)
\end{equation}
where, $\mathcal{T}_{1,I}^\text{L}(x_0)$ (or $\mathcal{T}_{1,II}^\text{L}(x_0)$) denotes the MFPT if the particle starts in region $I$ (or $II$).
 Applying these conditions determines the MFPT as a function of $x_0$. For $x_0=0$, it takes the following form
\begin{equation}\label{mfpt analytics linear}
\begin{split}
    \mathcal{T}^\text{L}_1=-&\frac{k_BT}{\Delta U^2}\Big(\frac{1}{k_1^2}+\frac{1}{k_2^2}\Big)-\frac{a}{k_1 \Delta U}-\frac{1-a}{k_2\Delta U}+\frac{k_BT}{k_1k_2\Delta U^2}\\
    &+\frac{k_BT}{\Delta U^2}\Big(\frac{1}{k_1^2}-\frac{1}{k_1k_2}\Big)e^{ak_1\Delta U/k_BT}+\frac{k_BT}{k_1k_2\Delta U^2}e^{\Delta U/k_BT}\\
    & +\frac{k_BT}{\Delta U^2}\Big(\frac{1}{k_2^2}-\frac{1}{k_1k_2}\Big)e^{k_2(1-a)\Delta U/k_BT}
    \end{split}
\end{equation}
At fixed values of $a$ and $\Delta U$, MFPT only depends on $k_1$ and $k_2$. However, due to the constraint in Eq.(\ref{eq:fbc sk}), only one of $k_1$ and $k_2$ is free. We take $k_1$ to be the free parameter so that $k_2$ is automatically determined by $k_2=(1-k_1a)/(1-a)$. In Fig.\ref{fig:single kink linear and non linear}(b), we plot $\mathcal{T}_1^{\text{L}}/\mathcal{T}_0^{\text{L}}$ as a function of $k_1$ for fixed $a=0.3$. Quite interestingly, for a range of $k_1$ values, this ratio remains below unity, implying that the modification leads to a reduction in the MFPT compared to the unmodified case in this parameter regime. This reduction can be attributed to the interplay of the times taken in traversing the two segments of the modified linear potential \cite{palyulin2012finite}. By tuning the strengths $k_1$ and $k_2$, the additional time required to cross the steeper segment—relative to crossing the same spatial portion of unmodified potential with strength $\Delta U$—can be outweighed by the time gained on the shallower segment. For appropriate choices of $k_1$ and $k_2$, this trade-off leads to a net decrease in the MFPT compared to the unmodified case. To quantify this, we introduce the factor speed-up which will be discussed next.

\textit{Maximal speedup for the process completion:}
We define the speedup factor as the ratio of the MFPT corresponding to the original unmodified potential to that corresponding to the potential modified by the introduction of intermediate barrier(s).
 In the present case, it is given by $\mathcal{T}_0^\text{L}/\mathcal{T}_1^\text{L}$. A ratio less than unity indicates that the modification results in slowing down the process, whereas a ratio greater than unity signifies the acceleration of the process.  In Fig.\ref{fig:single kink linear and non linear}(b), we plot the inverse of this factor with $k_1$ for a given $a$, which shows that for a range of $k_1$ values, the inverse speedup factor $\mathcal{T}_1^\text{L}/\mathcal{T}_0^\text{L}<1$, implying that in this range the first passage process is accelerated  by the modified potential, as compared to the unmodified one.

Further, as $k_1$ is varied, the inverse speed-up factor $\mathcal{T}_1^\text{L}/\mathcal{T}_0^\text{L}$ exhibits a minimum at a specific value, denoted by $k_1^*$.
We denote this minimum MFPT by $\mathcal{T}_1^{\text{L}^*}=\mathcal{T}_1^{\text{L}}|_{k_1=k_1^*}$, so that the speed-up ratio $\mathcal{T}_0^{\text{L}}/\mathcal{T}_1^{\text{L}^*}$ is maximum. The inverse maximal speedup factor is plotted as a function of $a$ in Fig.\ref{fig:single kink minimal MFPT vs a}(b). 
 For $a < 1/2$, the optimal strengths satisfy $k_1^* > k_2^*$, whereas for $a > 1/2$, $k_1^* < k_2^*$, indicating that in the optimal configuration, the region with the steeper strength occupies the smaller fraction of the spatial domain. Interestingly, $\mathcal{T}_1^{\text{L}^*}/\mathcal{T}_0^{\text{L}} \leq 1$, with equality only at $a = 1/2$. This implies that the minimum MFPT is strictly smaller than the MFPT of the unmodified linear potential, or in other words, the maximal speed-up ratio is greater than $1$ for all $a \neq 1/2$. At $a = 1/2$, the equality $\mathcal{T}_1^{\text{L}^*} = \mathcal{T}_0^{\text{L}}$ arises because the optimal strengths satisfy $k_1^* = k_2^*$ (see Fig.~\ref{fig:single kink minimal MFPT vs a}(b)), indicating that the optimal potential coincides with the original linear potential. These observations demonstrate that introducing an intermediate barrier into a linear potential enables a robust reduction in MFPT and therefore a robust speeding-up of the process completion across the entire range of $a$.
 \begin{figure*}
    \centering
    \includegraphics[width=1\linewidth]{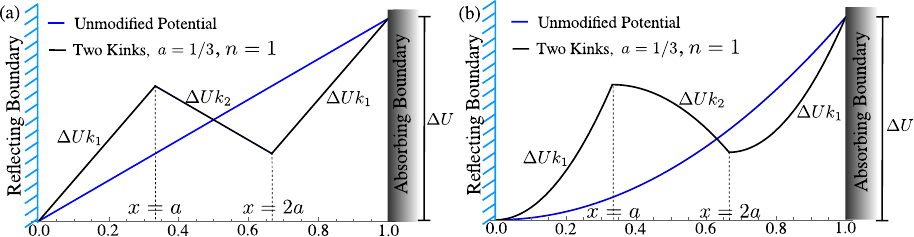}
    \caption{The figure presents the linear and harmonic potentials (in blue) along with their corresponding modified potentials after introducing $N = 2n = 2$ kinks (in black), as shown in panels (a) and (b), respectively. The modified potential consists of segments with alternating strengths $k_1$ and $k_2 $. In both cases, the kinks are equally spaced and positioned at $ x = a $ and $ x = 2a $, where $ a = 1/3$.}
    \label{fig:schematics multiple kinks L H}
\end{figure*}

\subsection{MFPT reduction in harmonic potential}
\label{sec:single kink harmonic potential}
We now turn our attention to the motion in the presence of a harmonic potential described by
\begin{equation}
 \label{eq:no kink non-linear potential}
 \frac{U^\text{H}_0(x)}{\Delta U}=x^2, \qquad \qquad 0\leq x\leq 1  . 
\end{equation}
The corresponding MFPT for the particle, starting at $x_0=0$, can be obtained by solving Eq.(\ref{mfpt differential equation}) with the two boundary conditions, and reads \cite{abkenar2017dissociation}
\begin{equation}
    \mathcal{T}_0^\text{H}=\frac{1}{2k_BT}\text{ }{ } _2F_2\Big(1,1;\frac{3}{2},2;\frac{\Delta U}{k_BT}\Big)
\end{equation}
where ${}_pF_q(a_1, \cdots, a_p;b_1, \cdots, b_q;z)$ stands for the generalized hypergeometric function.
We now introduce a kink in the harmonic potential at 
$x=a$ (see Fig.\ref{fig:single kink linear and non linear}(c)) defining a piecewise harmonic potential as follows 
\begin{equation}
\label{non linear potential}
\frac{U^\text{H}_1(x)}{\Delta U}=
    \begin{cases}
     k_1 x^2, \qquad \qquad \qquad \qquad 0\leq x\leq a \\
     k_2(x-a)^2+k_1 a^2 \qquad \quad a\leq x\leq 1
    \end{cases}
\end{equation}
where, $k_1$ and $k_2$ denote the strength of the potential segments which can take any real value.
The set up constraint naturally implies 
\begin{equation}
\label{eq:constraint harmonic sk}
    U_1^\text{H}(x=1)=\Delta U=\Delta U[k_1 a^2+k_2(1-a)^2].
\end{equation}
The associated MFPT for this modified potential is derived following Section~\ref{sec:single kink linear potential}, yielding 
\begin{equation}
\label{eq:harmonic_single_kink_analytical_expression}
\begin{split}
\mathcal{T}^\text{H}_1&= \frac{e^{\frac{a^2 k_1 \Delta U}{k_BT}} \pi \, {erf}\left( \frac{a \sqrt{k_1\Delta U}}{\sqrt{k_BT}} \right) \, {erf}i\left( \frac{(1 - a) \sqrt{k_2\Delta U}}{\sqrt{k_BT}} \right)}{4 \sqrt{k_1k_2} \Delta U} \\
& - \frac{\pi \, {erf}\left( \frac{(1 - a) \sqrt{k_2\Delta U}}{\sqrt{k_BT}} \right) \, {erf}i\left( \frac{(-1 + a) \sqrt{k_2\Delta U}}{\sqrt{k_BT}} \right)}{4 k_2 \Delta U} \\
&  +\frac{a^2}{2 k_BT} \, {}_2F_2\left( 1, 1; \frac{3}{2}, 2 ;\frac{a^2 k_1 \Delta U}{k_BT} \right)  \\
& -\frac{(-1 + a)^2}{2 k_BT}  \,{}_2F_2 \left(1, 1; \frac{3}{2}, 2; - \beta \right),
\end{split}
\end{equation}
where $\beta=(1-a)^2k_2\Delta U/k_BT$ and $erf(z)=\frac{2}{\sqrt{\pi}}\int_0^z e^{-y^2}dy$ denotes the error function, $erfi(z)=-i\,\text{erf}(iz)$ denotes the imaginary error function. Following a similar approach as in Section~\ref{sec:single kink linear potential}, we obtain Fig.\ref{fig:single kink linear and non linear}(d), which illustrates the dependence of the inverse speed-up factor, $\mathcal{T}_1^{\text{H}}/\mathcal{T}_0^\text{H}$, on $k_1$ for the harmonic case. As in the linear case, $\mathcal{T}_1^{\mathrm{H}} / \mathcal{T}_0^{\mathrm{H}} < 1$ over a range of $k_1$, indicating a reduced MFPT for the modified potential compared to the unmodified harmonic case. 

\textit{Maximal speedup for the process completion: } Likewise the piecewise linear potential, the inverse speed-up factor exhibits a minimum as shown in Fig.\ref{fig:single kink linear and non linear}(d),  which corresponds to the maximal speedup of the process completion. Further, Fig.\ref{fig:single kink minimal MFPT vs a}(d) presents the variation of the inverse maximal speed-up factor, $\mathcal{T}_1^{\text{H}^*} / \mathcal{T}_0^{\text{H}}$, and optimal strengths $k_1^*$ and $k_2^*$ with $a$, where $k_1^*$ (and $k_2^*$) for a given $a$ corresponds to the minimum MFPT and therefore the maximal speedup in the process completion for that specific value of $a$. See Fig.\ref{fig:single kink minimal MFPT vs a}(c) for a few representative optimal potential profiles. Further, Fig.\ref{fig:single kink minimal MFPT vs a}(d) shows that there exists a range of the parameter $a$ such that $k_1^* > k_2^*$, and at other values, $k_2^* > k_1^*$. Interestingly, we find $\mathcal{T}_1^{\text{H}^*}/\mathcal{T}_0^\text{H}<1$ for all $a$, demonstrating a robust reduction in the optimal MFPT across the entire range of $a$, thus showcasing a similar optimization.

\section{Introducing multiple intermediate barriers}
\label{sec:Multiple kinks}
Previously, we analyzed the effect of incorporating a single intermediate barrier on the controlled reduction of the mean first-passage time (MFPT). The aim of this section is to extend this analysis by introducing multiple intermediate barriers in the potential structure. As a means to achieve this, the potential is restructured into segments with alternating strengths of $k_1$ and $k_2$, starting with $k_1$; an illustration for a three-segment case is provided in Fig.~\ref{fig:schematics multiple kinks L H}. This introduces intermediate barriers, with their number $N$ equal to one less than the number of segments, $N+1$. For simplicity, we assume the kinks are uniformly spaced in the interval $x \in [0,1]$, located at $x = a, 2a, \dots, Na$ with $a = 1/(N+1)$, where $N$ is the number of kinks.

\subsection{Observation I: Intermediate barriers resulting from even number of segmentations \uline{do not} facilitate the further reduction of the MFPT with increasing segments}
\label{sec:even number of segments}
Let us 
consider the linear potential as in Eq.(\ref{eq:no kink linear potential}), partitioned now into 
$N+1$ segments by introducing $N$ kinks, where 
$N$ is assumed to be odd.
The potential then consists of $(N+1)/2$ segments each with strengths $k_1$ and $k_2$. Under this condition, the set up constraint, $U^\text{L}_N(x=1) = \Delta U$, implies:
\begin{equation}
    \label{eq:multiple odd kinks constraint linear}
        U^\text{L}_N(x=1)=\Delta U=\Delta U\frac{N+1}{2}k_1a+\Delta U \frac{N+1}{2}k_2a
\end{equation}

We analytically solve Eq.(\ref{mfpt differential equation}) to compute the MFPT, $\mathcal{T}^\text{L}_N$, using mathematica. This computation involves the two boundary conditions and $2N$ connecting conditions, two per kink (see Appendix \ref{app:fifth} for the details). Following the method in Section~\ref{sec:single kink}, we determine the minimum MFPT, $\mathcal{T}_N^{\text{L}^*}$, for various odd values of $N$ and compute the inverse maximal speed-up factor, $\mathcal{T}_N^{\text{L}^*}/\mathcal{T}_0^\text{L}$.  Fig.\ref{fig:multiple kinks linear results}(a) shows its variation along with the corresponding optimal strengths $k_1^*$ and $k_2^*$ as a function of odd $N$. The result $\mathcal{T}_N^{\text{L}^*}/\mathcal{T}_0^\text{L} = 1$ for all odd $N$ shows that introducing such intermediate barriers does not reduce the MFPT in the linear case.

We now analyse the case of harmonic potential. Similar to the case of linear potential, the constraint for $N+1$ harmonic segments($N$ odd) is
\begin{equation}
    \label{eq:multiple odd kinks constraint harmonic}
        U^\text{H}_N(x=1)=\Delta U=\Delta U\frac{N+1}{2}k_1a^2+\Delta U \frac{N+1}{2}k_2a^2
\end{equation}

Following the same procedure as for the linear case, we compute the inverse maximal speed-up factor, $\mathcal{T}_N^{\text{H}^*}/\mathcal{T}_0^\text{H}$, and the corresponding optimal strengths $k_1^*$ and $k_2^*$. Figure~\ref{fig:multiple kinks harmonic results}(a) shows their dependence on odd $N$. While the modified harmonic potential yields a lower MFPT than the unmodified case, the minimal MFPT increases with $N$, indicating that the maximum reduction occurs at $N = 1$. Thus, adding more kinks does not further accelerate the absorption.

\subsection{Observation II: Intermediate barriers resulting from odd number of segmentations of potential landscapes \uline{do} facilitate the successive reduction of the MFPT}
\label{sec:odd number of segments in linear potential}

\begin{figure}[t]
    \centering
    \includegraphics[width=1\linewidth]{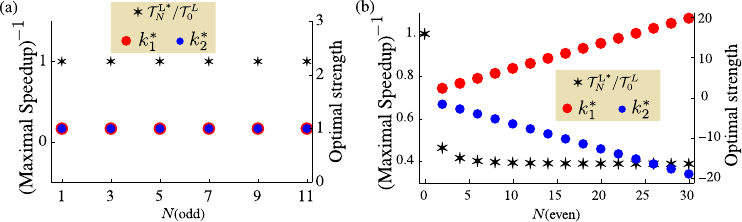}
    \caption{
      Variation of the inverse maximal speed-up factor, $\mathcal{T}_N^{\text{L}^*}/\mathcal{T}_0^\text{L}$, and the optimal strengths, with the number of kinks $N$. The left vertical axis represents the inverse maximal speed-up factor, while the right vertical axis corresponds to optimal strengths, $k_1^*$ and $k_2^*$. The barrier height is fixed at $\Delta U=5$ and $k_BT=1$. To obtain these results, the MFPT for a given $N$ is first computed analytically using Mathematica and the corresponding optimal MFPT is evaluated by numerical minimization. Panel
 (a) shows that for any odd $N$, the inverse maximal speed-up factor is unity, indicating no reduction in the MFPT. In these cases, the optimal strengths satisfy $k_1^* = k_2^*$. Panel (b) illustrates the behavior for even values of $N = 2n$. The inverse maximal speed-up factor is less than $1$ and decreases monotonically, indicating successive reduction of the MFPT with increasing $N$. The case $N = 0$ corresponds to the unmodified linear potential.
 }
    \label{fig:multiple kinks linear results}
\end{figure}

Let us now modify the linear potential as in Eq.(\ref{eq:no kink linear potential}) by dividing it into three segments (two kinks) as shown in Fig. \ref{fig:schematics multiple kinks L H}(a).
The resulting piecewise-linear potential is given by  
\begin{equation}
\label{eq:linear potential two kinks}
   \frac{U^\text{L}_2(x)}{\Delta U}=\begin{cases}
        k_1 x \hspace{3.5cm} 0\leq x\leq a\\
        k_2(x-a)+k_1a \hspace{1.7cm} a\leq x\leq 2a\\
        k_1(x-2a)+(k_1+k_2)a \hspace{0.45cm} 2a\leq x\leq 1
    \end{cases}
\end{equation}
Here, the kinks are located at $x=a$ and $x=2a$, with $a=1/3$. Imposing the constraint $U_2^\text{L}(x=1)=\Delta U$ yields $\Delta U=\Delta U[2k_1+k_2]a$. 
Generalizing to $N$ kinks (with even $N$), the potential is divided into $N+1$ segments. Letting $n = N/2$, the condition $U^\text{L}_N(x = 1) = \Delta U$ gives:
\begin{equation}
\label{eq:constraint MK linear}
   \Delta U=\Delta U(n+1)k_1a+\Delta U nk_2a
\end{equation}

The inverse maximal speed-up factor, $\mathcal{T}_N^{\text{L}^*}/\mathcal{T}_0^\text{L}$, along with the corresponding optimal strengths $k_1^*$ and $k_2^*$ for various even values of $N$, is computed following the procedure outlined in Section~\ref{sec:even number of segments}. As shown in Fig.~\ref{fig:multiple kinks linear results}(b), the inverse maximal speed-up factor decreases with increasing $N$, satisfying $\mathcal{T}_N^{\text{L}^*}/\mathcal{T}_0^\text{L} < 1$ for all $N > 0$ (with $N=0$ corresponding to the unmodified linear potential). This monotonic decrease indicates that adding more such intermediate barriers is beneficial.

\begin{figure}
    \centering
    \includegraphics[width=1\linewidth]{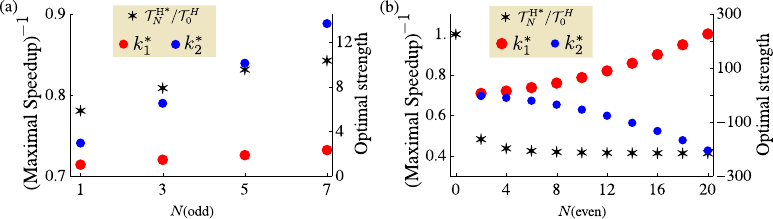}
    \caption{Variation of the inverse maximal speed-up factor, $\mathcal{T}_N^{\text{H}^*}/\mathcal{T}_0^\text{H}$, and the optimal strengths with $N$. The barrier height is fixed at $\Delta U=5$ and $k_BT=1$. 
Panel (a): For odd $N$, the inverse speed-up factor is less than one, indicating reduced MFPT as compared to that for unmodified potential. However, additional kinks doesn't lead to further reduction. Panel (b) illustrates the behavior for even values of $N = 2n$, where the introduction of kinks leads to a reduction in the optimal MFPT. Moreover, the minimal MFPT decreases monotonically with increasing $N$, indicating that the introduction of additional kinks further reduces the MFPT. The case $N = 0$ corresponds to the unmodified linear potential. 
    \label{fig:multiple kinks harmonic results}}
\end{figure}

We now analyze the case of harmonic potential.
As an example, consider partitioning the harmonic potential in Eq.~(\ref{eq:no kink non-linear potential}) into three segments, by introducing two kinks. The resulting piecewise-harmonic potential is given by
\begin{equation}
\label{eq:harmonic potential two kinks}
    \frac{U^\text{H}_2(x)}{\Delta U}=\begin{cases}
        k_1 x^2 \hspace{3.6cm} 0\leq x\leq a\\
       k_2(x-a)^2+k_1a^2 \hspace{1.6cm} a\leq x\leq 2a\\
        k_1(x-2a)^2+(k_1+k_2)a^2 \hspace{0.35cm} 2a\leq x\leq 1
    \end{cases}
\end{equation}
See Fig. \ref{fig:schematics multiple kinks L H}(b) for a schematic of this potential.
Imposing $U^\text{H}_2(x=1)=\Delta U$ implies $\Delta U=\Delta U(2k_1+k_2)a^2$. For any $N$(even) number of kinks, the constraint becomes 
\begin{equation}
    \label{eq:multiple even kinks constraint harmonic}
    U^\text{H}_N(x=1)=\Delta U=\Delta U(n+1)k_1a^2+\Delta U nk_2a^2
\end{equation}
where, $n=N/2$. A similar analysis to that for the linear case reveals that the inverse maximal speed-up factor, $\mathcal{T}_N^{\text{H}^*}/\mathcal{T}_0^\text{H}$,  decreases monotonically with increasing number of such intermediate barriers, or equilvalently, with increasing even $N$(see Fig.\ref{fig:multiple kinks harmonic results}(b)). The figure also shows the variation of $k_1^*$ and $k_2^*$ with $N$.

We have thus demonstrated that the minimum MFPT decreases monotonically with increasing even values of $N$ in both the linear and harmonic cases. This monotonic decrease implies that the maximal speedup achievable in the completion time of the process increases with increasing even $N$, i.e., $\mathcal{T}_0^\text{L}/\mathcal{T}_{N+2}^{\text{L}^*}>\mathcal{T}_0^\text{L}/\mathcal{T}_N^{\text{L}^*}$ (similarly $\mathcal{T}_0^\text{H}/\mathcal{T}_{N+2}^{\text{H}^*}>\mathcal{T}_0^\text{H}/\mathcal{T}_N^{\text{H}^*}$ for the harmonic case) for all even $N$. This behavior is attributed to ``additional degrees of freedom" which emanates from the incorporation of more kinks/barriers into the potential profile. Each pair of additional kinks (i.e., increasing $N$ to $N+2$) seems to introduce greater control in shaping the potential profile as the enhanced degree redistributes the strengths of the uphill and downhill segments in a manner that reduces the overall traversal time, resulting in $\mathcal{T}_{N+2}^{\text{L}^*} < \mathcal{T}_N^{\text{L}^*}$ (and $\mathcal{T}_{N+2}^{\text{H}^*} < \mathcal{T}_N^{\text{H}^*}$).

\begin{figure}
    \centering
    \includegraphics[width=1\linewidth]{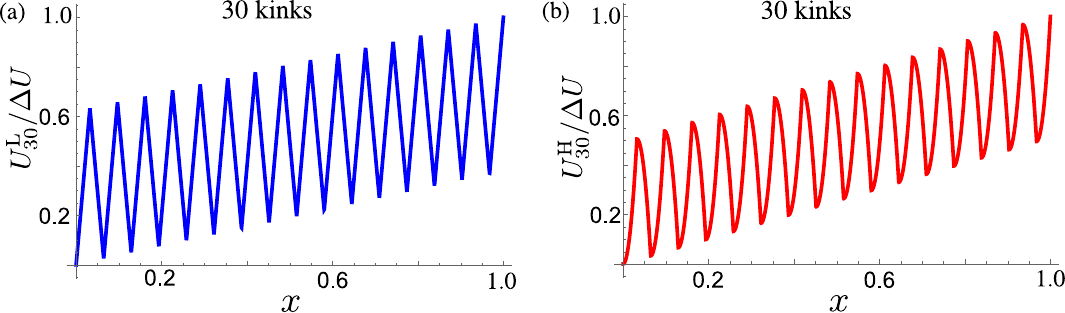}
    \caption{Panels (a) $\&$ (b): Optimally restructured potential configurations in linear and harmonic case for $N=30$, $\Delta U=5$ and $k_BT=1$.}
    \label{fig:optimal profiles}
\end{figure}

We continue to investigate the behavior of the optimal potential profiles in the large and even $N \to \infty$ limit. To this end, we plotted the optimal profiles for large $N (= 30)$ in Fig.~\ref{fig:optimal profiles}, which is helpful in understanding the asymptotic structure. To determine the scaling of the optimal strength $k_1^*$, we fitted the dataset $\{k_1^*, N\}$ for even values of $N$ from $N = 2$ to $N = 20$, and confirmed that it correctly gives the result at larger values of $N$ (up to $N = 100$). For $\Delta U = 5$, the resulting fit for the linear case is
\begin{equation}
\label{Eq:fit linear}
    k_1^* \approx 1.058 + 0.618\, N,
\end{equation}
while for the harmonic case it is
\begin{equation}
\label{Eq:fit harmonic}
    k_1^* \approx 1.185 + 1.505\, N + 0.487\, N^2,
\end{equation}
where the symbol $\approx$ indicates that the coefficients have been rounded to three decimal places. Once $k_1^*$ is known, the corresponding value of $k_2^*$ can be computed using the constraint equations, Eq.~(\ref{eq:constraint MK linear}) for the linear case and Eq.~(\ref{eq:multiple even kinks constraint harmonic}) for the harmonic case.

\section{Conclusions and Discussion}
\label{sec:conclusion}
\begin{table*}
\caption{Summary of main results. We recall that $\mathcal{T}^\text{L} _N$ denotes the MFPT for a piecewise linear potential with $N+1$ segments or $N$ kinks. The case $N=0$ refers to the pure linear potential. The MFPT for the piecewise Harmonic potential with $N$ kinks is $\mathcal{T}^\text{H} _N$, where again, the case of $N=0$ refers to the pure Harmonic potential. }
\label{tab:summary}
\centering
\renewcommand{\arraystretch}{1.4}
\newcolumntype{C}[1]{>{\centering\arraybackslash}p{#1}} 
\begin{tabular}{|l|c|l|l|}
\hline
\textbf{Potential Type} & \textbf{\# of kinks ($N$)} & \textbf{Optimal Shape} & \textbf{MFPT Behavior with $N$} \\
\hline
\multirow{2}{*}{Piecewise Linear} 
  & Even & Non-uniform segment strengths & $ \mathcal{T}^{\text{L}^*} _N<\mathcal{T}_0^\text{L}$ and $\mathcal{T}_N^{\text{L}^*}$ monotonically decreases with $N$  \\
  & Odd  & Uniform linear potential  &  Linear case always optimal \\
\hline
\multirow{2}{*}{Piecewise Harmonic} 
  & Even & Non-uniform segment strengths  & $ \mathcal{T}^{\text{H}^*} _N<\mathcal{T}_0^\text{H}$ and $\mathcal{T}_N^{\text{H}^*}$ monotonically decreases with $N$ \\
  & Odd  & Non-uniform segment strengths  &  $\mathcal{T}^{\text{H}^*} _N<\mathcal{T}_0^\text{H}$ for some $N$, but $\mathcal{T}_N^{\text{H}^*}$ increases with $N$  \\
\hline
\end{tabular}
\end{table*}

In this work, we investigate how the mean first-passage time (MFPT) of a Brownian particle can be controlled by tailoring the shape of the external potential. We show that the MFPT required for a particle to cross a potential barrier of height $\Delta U$ can be markedly reduced by systematically modifying the potential landscape, without altering the overall barrier height. In particular, we demonstrate that replacing a single large barrier with a sequence of smaller intermediate barriers—constructed by partitioning the landscape into segments via the introduction of kinks—leads to a significant acceleration of the first-passage process.

These observations were based on rigorous analysis of the MFPT in a linear and harmonic potential, both unperturbed and modified likewise. We began our analysis by demonstrating that the introduction of a single intermediate barrier in the potential can result in the reduction of the MFPT compared to that of the original unmodified potential. Building on this, we extended the study to the case of multiple equidistant kinks, denoted by $N$, and investigated the corresponding effects on the MFPT. We further examined a speed-up factor to quantitatively account for the efficacy of the transition rendered by the intermediate barriers. The results are summarized in Table~\ref{tab:summary}.

Similar to the MFPT, another important observable in the study of first-passage processes is the variance of the first-passage time, which typically characterizes the fluctuations around the MFPT. While the MFPT is the average time required for a stochastic process to reach the target for the first time, the variance quantifies the deviation around the MFPT. Although a detailed investigation of the variance of the FPT lies beyond the objective of this work, we briefly examine it for a few cases. We first study the variance corresponding to linear potential (as in Eq.~\eqref{eq:no kink linear potential}), and that corresponding to modified linear potential with one intermediate barrier (see Eq.~\eqref{Linear potential}). We then follow this analysis for the case of harmonic potential  (as in Eq.~\eqref{eq:no kink non-linear potential}) and for modified harmonic potential with one intermediate barrier (see Eq.~\eqref{non linear potential}). Skipping details from the analysis presented in Appendix~\ref{sec:app sixth}, we can conclude that adding intermediate barrier can also reduce the fluctuations of the escape time of a Brownian particle. 

%

In summary, this study has demonstrated that modifying the potential landscape can be an effective strategy for reducing the mean first-passage time (MFPT)  and the variance of the first-passage time of an overdamped Brownian particle. In particular, the piecewise harmonic potential offers a realistic and experimentally accessible model. It can be readily implemented using arrays of optical traps \cite{grier2006holographic,roichman2007colloidal,ferrer2024experimental}, making it a practical framework for experimentalists and a promising tool for exploring such mechanisms in real systems.

This work also points out to several important directions that remain open. Although our findings suggest that introducing intermediate barriers into complex potential landscapes generally lowers the MFPT, a rigorous theoretical proof of this observation is certainly lacking. Establishing such a result or identifying the extent to which these controls are robust would be a crucial step forward, with potential implications for processes such as molecular or ionic transport through membranes and channels \cite{dagdug2024diffusion,thorneywork2020direct}. Another promising direction is to explore the optimal number of subdivisions in complex potential landscapes, extending similar ideas as in \cite{bekele1996optimal}, where the potential was partitioned into linear segments within a specific setup which preserves the area under the potential. Unlike our current work that achieves MFPT reduction by tuning the strengths of the potential segments, this optimization with regard to the number of barriers could be another prospecting robust control over the barrier crossing time. In addition, it would be worthwhile to investigate whether similar reductions in first-passage times can be achieved for inertial, underdamped Brownian particles through analogous modifications of the potential. Another interesting direction involves examining the effect of intermediate barriers in anomalous diffusive systems. Indeed, a study by Palyulin and Metzler \cite{palyulin2013speeding} demonstrated that the first-passage dynamics of a subdiffusive continuous-time random walk can be accelerated by introducing a finite potential barrier. However, the influence of multiple intermediate barriers (as well as their effect to other non-diffusive processes such as \cite{capala2020levy}), as done in this paper, on the MFPT in such systems remains to be explored.

Another promising direction would be to investigate whether introducing intermediate barriers can accelerate the mean first-passage time (MFPT) in stochastic processes with multiplicative noise, such as geometric Brownian motion \cite{stojkoski2021geometric}.
Finally, it would be interesting to explore the combined optimization effects arising from the interplay between intermediate barriers and stochastic resetting. However, one may need to adapt an annealed
resetting algorithm \cite{sar2023resetting} to see such effect in confined geometries such as ion channels \cite{dagdug2024diffusion,jain2023fick,pal2024channel}.

\section{Acknowledgment}
The numerical calculations reported in
this work were carried out on the Kamet cluster, which is maintained and supported by the Institute of Mathematical Science’s High-Performance Computing Center. 
VK and AP gratefully acknowledge research support from the Department of Atomic Energy, Government of India via the Apex Projects. AP acknowledges the support from the International Research Project (IRP) titled ``Classical and quantum dynamics in out of equilibrium systems'' by CNRS, France. OS and AP thank the Korea Institute for advanced Study (KIAS), Seoul, for hospitality during the conference ``Nonequilibrium Statistical Physics of Complex Systems''  where several discussions related to the project took place.

\appendix

\section{ Discussions on Connecting Conditions and MFPT Computation in the Presence of Intermediate Barriers in Linear and Harmonic Potentials} 
\label{app:fifth}
The aim of this appendix is to derive the matching conditions, for the case when a potential has a single and multiple kinks in it. Although such conditions are natural and can generically be found in textbooks such as \cite{gardiner2009stochastic,risken1996fokker}, we provide this discussion for completeness. Consider a generic potential $U(x_0)$ which has a kink at $x_0=a$. Therefore, $U'(x_0)$ is finitely discontinuous and $U''(x_0)$ is infinitely discontinuities at $x_0=a$. For such potentials, these discontinuities appear in the left hand side (LHS) of the differential equation, Eq.(\ref{mfpt differential equation}). However, the right hand side (RHS) of Eq.(\ref{mfpt differential equation}) is identically equal to $-1$ and hence remains continuous, regardless of the form of the potential $U(x_0)$ on the LHS. It follows that the discontinuities on the LHS must cancel in such a way that the LHS becomes continuous, consistent with the continuity of the RHS.

There are two potential scenarios for the cancellation of this discontinuity
\begin{enumerate}
    \item  $\mathcal{T}'(x_0)$ is finitely discontinuous at $x_0=a$ resulting in $\mathcal{T}''(x_0)$ being infinitely discontinuous there. 
    \item $\mathcal{T}'(x_0)$ is continuous at $x_0=a$ while $\mathcal{T}''(x_0)$ is finitely discontinuous at that point.
\end{enumerate}
 The first possibility is rejected as an infinite discontinuity cannot cancel out a finite discontinuity. Therefore, we are left only with the second option. This implies that $\mathcal{T}''(x_0)$ must be finitely discontinuous at $x_0=a$, which can get cancelled by the finite discontinuity of $U'(x_0)$ at $x_0=a$. It therefore follows that $\mathcal{T}'(x_0)$ must remain continuous at $x_0=a$. This automatically implies that $\mathcal{T}(x_0)$ must also be continuous at this point. Thus, we have found the two connecting conditions : $\mathcal{T}(x_0)$ and $\mathcal{T}'(x_0)$ must be continuous at $x_0=a$.
 
 Note that the two connecting conditions can also be derived as follows. 
First, we recast Eq.(\ref{mfpt differential equation}) 
into the following form
\begin{equation}
\label{eq: mfpt differential eqn 2nd form}
    k_BT \frac{d}{dx_0} \left[ e^{-U(x_0)/k_BT} 
    \frac{d}{dx_0} \mathcal{T}(x_0) \right] 
    = -e^{-U(x_0)/k_BT}.
\end{equation}

Since $a$ is the position of the kink, we integrate 
Eq.(\ref{eq: mfpt differential eqn 2nd form}) from 
$a - \epsilon$ to $a + \epsilon$, where $\epsilon$ is a small 
positive number. The right-hand side becomes zero because 
$e^{-U(x_0)/k_BT}$ is a continuous function, as $U(x_0)$ 
is continuous. Therefore, we obtain:
\begin{equation}
    e^{-U(x_0)/k_BT} \frac{d}{dx_0} \mathcal{T}(x_0) 
    \Big|_{x_0 = a - \epsilon}^{x_0 = a + \epsilon} = 0.
\end{equation}

Since $e^{-U(x_0)/k_BT}$ is continuous at $x_0 = a$, the above 
equation implies that
\begin{equation}
    \frac{d}{dx_0} \mathcal{T}(x_0 = a - \epsilon) 
    = \frac{d}{dx_0} \mathcal{T}(x_0 = a + \epsilon).
\end{equation}
Writing it in terms of the notations $\mathcal{T}_{1,I}^\text{L}$ and $\mathcal{T}_{1,II}^\text{L}$  used in Section \ref{sec:single kink linear potential}, implies Eq.(\ref{eq: connecting condition 1}). Since the derivative of the function MFPT is continuous, 
it follows that the MFPT itself must also be continuous at $x_0=a$. 
This yields Eqs. (\ref{eq: connecting condition 1}) and (\ref{eq: connecting condition 2}) of the main text.

The derivations of the connecting conditions, given in Eqs.(\ref{eq: connecting condition 1}) and (\ref{eq: connecting condition 2}), 
do not rely on any specific functional form of the potential $U(x_0)$. The only requirement for the validity of these conditions is that the potential 
$U(x_0)$ is continuous at $x_0=a$, the position of the kink. Consequently, the same connecting 
conditions apply to the MFPT corresponding to the 
piecewise-harmonic potential defined in Eq.(\ref{non linear potential}).

To compute the MFPT for the case of multiple intermediate barriers in a linear or harmonic potential, it is important to note that there are two connecting conditions at the position of each kink. In addition to these, we already have two boundary conditions: a reflecting boundary at $x_0 = 0$ and an absorbing boundary at $x_0 = 1$. Thus, for $N$ kinks, we have a total of $2N + 2$ conditions that the MFPT function $\mathcal{T}(x_0)$ must satisfy.
Furthermore, the presence of $N$ kinks implies that the domain is divided into $N + 1$ segments or regions. The solution of the differential equation (Eq.\eqref{mfpt differential equation}) within each region contains two undetermined integration constants. Therefore, across all $N + 1$ regions, there are $2(N + 1) $ integration constants in total.
Since the number of independent conditions matches the number of unknown constants, all integration constants are uniquely determined. Consequently, the corresponding MFPT is fully specified.

\section{Computation of variance in first-passage time for the linear and harmonic potentials with single intermediate barrier}
\label{sec:app sixth}
In this section, we analyze the variance of the first-passage time (FPT) for a Brownian particle in the said potential configurations. We first consider the case of a linear potential, Eq.~\eqref{eq:no kink linear potential}, and compare it to a modified linear potential that includes a single intermediate barrier, Eq.~\eqref{Linear potential}, while keeping the overall barrier height $\Delta U$ identical in both cases. Remarkably, we find that analogous to the behavior observed for the MFPT, the variance of the FPT in the modifed linear potential can be reduced relative to that in the purely linear potential (see Fig.~\ref{fig:sec moment}(a)). Then, we extend this analysis to the harmonic potential (Eq.~\eqref{eq:no kink non-linear potential}) and a modified harmonic potential with one intermediate barrier, Eq.~\eqref{non linear potential}. A similar trend is observed in this case as well, as illustrated in Fig.~\ref{fig:sec moment}(b).

To delve deeper, we start by denoting the second moment of the FPT by $\langle t_\text{{FP}}^2\rangle$. Accordingly, its variance is expressed by 
 \begin{equation}
     \text{Var(}t_\text{FP})=\langle t_\text{{FP}}^2\rangle-\mathcal{T}^2.
 \end{equation}
 The second moment satisfies the following differential equation obtained via the backward Fokker-Planck formalism\cite{gardiner2009stochastic},
\begin{equation}
\label{eq:secong moment general}
    2\mathcal{T}(x_0)=U'(x_0)\partial_{x_0}[\langle t_\text{{FP}}^2\rangle(x_0)]-k_BT \partial_{x_0}^2[\langle t_\text{{FP}}^2\rangle(x_0)],
\end{equation}
where, $\langle t_\text{{FP}}^2\rangle(x_0)$ denotes the second moment of the FPT for the particle starting its motion at $x_0$.

For the linear and harmonic potentials, Eq.~\eqref{eq:secong moment general} can be solved directly for $\langle t_{\mathrm{FP}}^2 \rangle(x_0)$, subject to appropriate boundary conditions (specified below). However, for the piecewise-linear and piecewise-harmonic potentials defined in Eqs.~\eqref{Linear potential} and \eqref{non linear potential}, respectively, the derivatives $U'(x_0)$ are discontinuous at the kink position $x_0 = a$. Consequently, Eq.~\eqref{eq:secong moment general} must be solved separately in two regions: region~$I$, $x_0 \in [0, a]$, and region~$II$, $x_0 \in [a, 1]$. The complete solution thus requires two connecting conditions at $x_0 = a$, in addition to the boundary conditions. The procedure outlined in Appendix~\ref{app:fifth} implies that the second moment $\langle t_{\mathrm{FP}}^2 \rangle(x_0)$ and its first derivative is continuous at $x_0=a$, despite a finite discontinuity in $U'(x_0)$ at $x_0=a$. (This conclusion requires the continuity of $\mathcal{T}(x_0)$, which has been explicitly established in Appendix~\ref{app:fifth}.)
This provides the necessary connecting conditions for determining the solutions.

\subsection{Reduction of the FPT variance in linear potential}
Here, we first determine the variance of the FPT corresponding to the linear potential in Eq.~\eqref{eq:no kink linear potential}. For this potential Eq.\eqref{eq:secong moment general} takes the form
\begin{equation}
\label{Eq:sec mom lin NK}
    2\mathcal{T}_0^\text{L}(x_0)=\Delta U\partial_{x_0}[\langle t_\text{{FP}}^2\rangle_0^\text{L}(x_0)]-k_BT \partial_{x_0}^2[\langle t_\text{{FP}}^2\rangle_0^\text{L}(x_0)]
\end{equation}
where $\langle t_{\mathrm{FP}}^{2}\rangle_0^{\text{L}}(x_0)$ denotes the second moment of the FPT for a particle starting at position $x_0$ and moving under the linear potential (no intermediate barrier) defined in Eq.~\eqref{eq:no kink linear potential}. Eq.~\eqref{Eq:sec mom lin NK} is solved subject to the reflecting and absorbing boundary conditions:
\begin{equation}
   \partial_{x_0} \langle t_\text{{FP}}^2\rangle_0^\text{L}(x_0=0)=0, \qquad \langle t_\text{{FP}}^2\rangle_0^\text{L}(x_0=1)=0.
\end{equation}
 Once $\langle t_{\mathrm{FP}}^2\rangle_0^{\mathrm{L}}(x_0)$ is obtained, the variance of the FPT follows from the relation
\begin{equation}
\text{Var}(t_\text{FP})_0^\text{L}(x_0)=\langle t_\text{{FP}}^2\rangle_0^\text{L}(x_0)-(\mathcal{T}_0^\text{L})^2(x_0),
\end{equation}
which, upon evaluation at $x_0=0$, yields the following expression for the variance of the FPT for a particle starting at $x_0=0$:
\begin{equation}
\label{Eq:variance linear NK}
    \begin{split}
        \text{Var}(t_\text{{FP}})_0^\text{L}(x_0=0)&\equiv  \text{Var}(t_\text{{FP}})_0^\text{L}\\
        & =\Bigg[\frac{4(k_BT)^2}{\Delta U^4}-\frac{4k_BT}{\Delta U^3}\Bigg]e^{\Delta U/k_BT}\\
        &+\frac{(k_BT)^2}{\Delta U^4}e^{2\Delta U/k_BT}-\frac{5(k_BT)^2}{\Delta U^4}-\frac{2k_BT}{\Delta U^3}
    \end{split}
\end{equation}

Next consider the piecewise linear potential $U_1^\text{L}$ in Eq. \eqref{Linear potential} (one intermediate barrier). The corresponding quantities in this case are denoted by $\text{Var}(t_{\mathrm{FP}})_1^{\mathrm{L}}(x_0)$ for the variance of the first-passage time and by $\langle t_{\mathrm{FP}}^2 \rangle_1^{\mathrm{L}}(x_0)$ for its second moment. These quantities are related through
\begin{equation}
\label{eq:sec mom var}
    \text{Var}(t_\text{FP})_1^\text{L}(x_0)=\langle t_\text{FP}^2\rangle_1^\text{L}(x_0)-(\mathcal{T}_1^\text{L})^2(x_0).
\end{equation}
For this potential in Eq.~\eqref{Linear potential}, Eq.~\eqref{eq:secong moment general} must be expressed separately in the two regions of the domain. The resulting equations take the form
\begin{equation}
\label{eq:secmom lin reg1 mfpt eqn}
    2\mathcal{T}_{1,I}^\text{L}(x_0)=\Delta Uk_1\partial_{x_0}[\langle t_\text{{FP}}^2\rangle_{1,I}^\text{L}(x_0)]-k_BT \partial_{x_0}^2[\langle t_\text{{FP}}^2\rangle_{1,I}^\text{L}(x_0)],
\end{equation}
for $x_0\in[0,a]$ and, 
\begin{equation}
\label{eq:secmom lin reg2 mfpt eqn}
\begin{split}
    2\mathcal{T}_{1,II}^\text{L}(x_0)=&\Delta Uk_2\partial_{x_0}[\langle t_\text{{FP}}^2\rangle_{1,II}^\text{L}(x_0)]\\
    &-k_BT \partial_{x_0}^2[\langle t_\text{{FP}}^2\rangle_{1,II}^\text{L}(x_0)], 
    \end{split}
\end{equation}
for $x_0\in[a,1]$.
Here, subscript $I$ (respectively $II$) denotes the quantities in region $I$ (respectively $II$).

Since the second moment and its first derivative is continuous at $x_0=a$, we have the following connecting conditions: 
\begin{equation}
\label{eq:secmom cont cond}
    \langle t_\text{FP}^2\rangle^\text{L}_{1,I}(x_0=a)=\langle t_\text{FP}^2\rangle^\text{L}_{1,II}(x_0=a)
\end{equation}
and
\begin{equation}
\label{eq:secmom der cond}
    \partial_{x_0}\langle t_\text{FP}^2\rangle^\text{L}_{1,I}(x_0=a)=\partial_{x_0}\langle t_\text{FP}^2\rangle^\text{L}_{1,II}(x_0=a).
\end{equation}
The reflecting boundary at $x_0=0$ implies that 
\begin{equation}
\label{eq:secmom ref bdy}
    \partial_{x_0}\langle t_\text{FP}^2\rangle^\text{L}_{1,I}(x_0=0)=0
\end{equation}
and the absorbing boundary leads to
\begin{equation}
\label{eq:secmom abs bdy}
    \langle t_\text{FP}^2\rangle^\text{L}_{1,II}(x_0=1)=0.
\end{equation}
We obtained analytical expressions for $\langle t_{\mathrm{FP}}^2\rangle_{1,I}^{\mathrm{L}}(x_0)$ and $\langle t_{\mathrm{FP}}^2\rangle_{1,II}^{\mathrm{L}}(x_0)$ (not shown here) by solving Eqs.~\eqref{eq:secmom lin reg1 mfpt eqn} and~\eqref{eq:secmom lin reg2 mfpt eqn}, subject to the connecting and the boundary conditions specified in Eqs.~\eqref{eq:secmom cont cond}, \eqref{eq:secmom der cond}, \eqref{eq:secmom ref bdy}, and~\eqref{eq:secmom abs bdy}. 
 The second moment of the first-passage time for a particle starting at $x_0 = 0$ is then given by $ \langle t_{\mathrm{FP}}^2 \rangle_{1,I}^{\mathrm{L}}(x_0 = 0)$. This quantity determines the corresponding variance of the first-passage time through Eq.~\eqref{eq:sec mom var}, evaluated at $x_0 = 0$, which we denote as $\text{Var}(t_{\mathrm{FP}})_1^\text{L}$ for brevity. The variance $\text{Var}(t_{\mathrm{FP}})_1^\text{L}$ depends on the system parameters $k_1$, $k_2$, $\Delta U$, and $a$.
 However, according to the fixed barrier height condition given in Eq.~\eqref{eq:fbc sk}, $k_2$ can be expressed as function of $k_1$ and $a$. Hence $\text{Var}(t_{\mathrm{FP}})_1^\text{L}$ is now a function of $k_1$, $\Delta U$ and $a$, which turns out to be
 \begin{equation}
\begin{split}
\text{Var}(t_{\mathrm{FP}})_1^\text{L} =\; & \frac{(k_BT)^2 e^{\frac{2 \Delta U}{k_BT}}}{k_1^2 k_2^2 \Delta U^4}
+ \frac{4 k_BT(k_BT- \Delta U \theta)}{k_1^2 k_2^2 \Delta U^4} e^{\frac{\Delta U}{k_BT}} \\
& + \frac{2 (k_BT)^2 \alpha}{k_1^2 k_2^3 \Delta U^4} e^{\frac{(2 - a k_1) \Delta U}{k_BT}}
- \frac{2 (k_BT)^2 \alpha}{k_1^3 k_2^2 \Delta U^4} e^{\frac{(1 + a k_1) \Delta U}{k_BT}} \\
& + \frac{(k_BT)^2 \alpha^2}{k_1^2 k_2^4 \Delta U^4} e^{\frac{2 (1 - a k_1) \Delta U}{k_BT}}
+ \frac{(k_BT)^2 \alpha^2}{k_1^4 k_2^2 \Delta U^4} e^{\frac{2 a k_1 \Delta U}{k_BT}} \\
& + \left( \frac{4 a k_BT\alpha}{k_1^3 k_2 \Delta U^3}
- \frac{2 (k_BT)^2 \beta}{k_1^4 k_2^3 \Delta U^4} \right) e^{\frac{a k_1 \Delta U}{k_BT}} \\
& + \left( \frac{4 a k_BT\alpha}{k_1 k_2^3 \Delta U^3}
- \frac{4 k_BT\alpha}{k_1 k_2^3 \Delta U^3}
- \frac{2 (k_BT)^2 \gamma}{k_1^3 k_2^4 \Delta U^4} \right)  \\
& \times e^{\frac{(1 - a k_1) \Delta U}{k_BT}}
+ \frac{(k_BT)^2 \delta}{k_1^4 k_2^4 \Delta U^4}
+ \frac{2 k_BT\phi}{k_1^3 k_2^3 \Delta U^3}
\end{split}
\end{equation}
where, $k_2=(1-k_1a)/(1-a)$, $\alpha=(k_1-k_2)$, $\beta=(k_1^3+k_1^2k_2-2k_2^3)$, $\gamma=(k_2^3-2k_1^3+k_1k_2^2)$, $\delta=(-5k_1^4+2k_1^3k_2+k_1^2k_2^2+2k_1k_2^3-5k_2^4)$, $\theta=(k_1-ak_1+ak_2)$ and $\phi=[(-1+a)k_1^3-ak_2^3]$.

For the purpose of analysis, we fix $k_BT=1$,  $\Delta U = 3$ and $a = 0.3$. This implies that $\text{Var}(t_\text{{FP}})_0^\text{L}$ is fully determined from Eq.~\eqref{Eq:variance linear NK}. However, variance $\text{Var}(t_{\mathrm{FP}})_1^\text{L}$ remains a function of $k_1$. For the sake of comparison, we plot the variation of the ratio $\text{Var}(t_{\mathrm{FP}})_1^\text{L}/\text{Var}(t_{\mathrm{FP}})_0^\text{L}$, which we call scaled variance, with $k_1$ in FIG.~\ref{fig:sec moment}(a). The plot shows that for a range of $k_1$ values, the scaled variance is less than unity, indicating that the introduction of an intermediate barrier within the original linear potential results in a reduction of the variance compared to that of the original process. 

\subsection{Reduction of the FPT variance in harmonic potential}

\begin{figure}
    \centering
    \includegraphics[width=\linewidth]{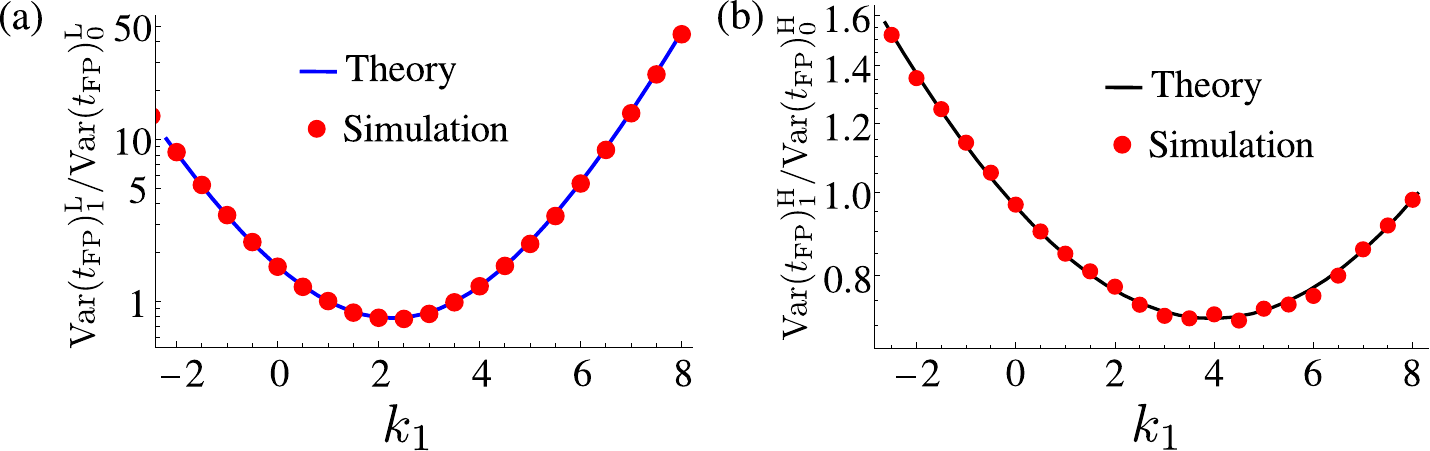}
    \caption{Semi-logarithmic plots of the scaled variance of the FPT for $k_BT=1$, $\Delta U = 3$ and $a = 0.3$, shown for (a) the piecewise linear potential (Eq.~\eqref{Linear potential}) and (b) the piecewise harmonic potential (Eq.~\eqref{non linear potential}). Solid lines represent the analytical solution of corresponding differential equations and the red dots correspond to the microscopic simulation of Eq.~\eqref{eq:Langevin eqn}. The curves falling below unity reflect a reduction in the FPT variance. }
    \label{fig:sec moment}
\end{figure}

Likewise the case of linear and modified linear potential with one intermediate barrier, we now analyze the variance of the harmonic, (Eq.\eqref{eq:no kink non-linear potential}), and modified harmonic potential with one intermediate barrier in Eq.~\eqref{non linear potential}.
Following the similar procedure as  in the precious subsection, we determine the scaled variance, $\text{Var}(t_{\mathrm{FP}})_1^\text{H}/\text{Var}(t_{\mathrm{FP}})_0^\text{H}$, by numerically solving the corresponding differential equations. Fixing the parameters at same values as in the case of linear and piecewise-linear potential, we plot the scaled variance as a function of $k_1$ in FIG.~\ref{fig:sec moment}(b). Like the previous case, the curve shows that for a range of $k_1$ values, scaled variance $\text{Var}(t_{\mathrm{FP}})_1^\text{H}/\text{Var}(t_{\mathrm{FP}})_0^\text{H}$ is less than unity. This indicates that introducing an intermediate barrier into a harmonic potential is indeed useful in reducing the FPT variance.

 \section{Simulation Specification}
 To generate the numerical data presented in Fig.\ref{fig:single kink linear and non linear} , we simulated $10^5$ independent trajectories of a Brownian particle. Each trajectory was initialized at $x = 0$ and evolved until the particle was absorbed at the boundary $x = 1$. The time evolution of the particle in each microscopic time scale $dt$ can be written in terms of a discretized form of Langevin equation in \ref{eq:Langevin eqn} using Euler–Maruyama approximation such as
\begin{equation}
    x(t + \Delta t) = x(t) - U'(x(t))\,\Delta t + \sqrt{2D\Delta t}\,\mathcal{N}(0,1),
\end{equation}
where $\mathcal{N}(0,1)$ denotes a Gaussian random variable with zero mean and unit variance. For the data corresponding to Fig.\ref{fig:single kink linear and non linear}(b), $dt=10^{-5}$ was used, while for Fig.\ref{fig:single kink linear and non linear}(d), $ dt = 10^{-7}$ was employed to ensure numerical accuracy. In both cases, the barrier height was fixed at $\Delta U = 3$, and the position of the kink was set to $a = 0.3$.

\twocolumngrid

\bibliography{KSUTET}

\end{document}